\documentclass[12pt,preprint]{aastex}
\usepackage{epsfig}
\usepackage{natbib}

\begin{document}
\voffset-0.5cm
\newcommand{\gsim}{\hbox{\rlap{$^>$}$_\sim$}}
\newcommand{\lsim}{\hbox{\rlap{$^<$}$_\sim$}}

\title{Analytical Expressions  For Light-Curves \\
Of Ordinary And Superluminous Supernovae Type Ia}

\author{Shlomo Dado\altaffilmark{1} and Arnon Dar\altaffilmark{1}}

\altaffiltext{1}{Physics Department, Technion, Haifa 32000, Israel}

\begin{abstract} 
Supernovae of type Ia (SNeIa) can be produced by the 
explosion of slowly-rotating carbon-oxygen white dwarfs whose mass 
increases beyond a critical value by mass accretion. Collision with 
circumstellar material during their photospheric and early nebular phase 
can enhance the bolometric luminosity of otherwise ordinary SNeIa to 
become superluminous. A few simplifying assumptions lead to a simple 
analytic master formula, which describes quite well the bolometric 
light-curves of ordinary SNeIa and supeluminous SNeIa in terms of few 
initial physical parameters. Other main properties of SNeIa, including the 
empirical 'brighter-slower' Phillips' relation that was used to 
standardize ordinary SNeIa as distance indicators and led to the discovery 
of the accelerating expansion of the universe, are reproduced. 
\end{abstract}


\maketitle

\section{Introduction}

Despite large observational, theoretical and numerical efforts over 
decades, supernovae explosions are not fully understood. Standard stellar 
evolution theory predicts that stars with initial mass ${\rm M\gsim 8\, 
M_\odot}$ end their short life in core-collapse supernovae explosions, 
while long-lived stars of mass ${\rm \lsim 8\, M_\odot}$ eject most of 
their outer layers, leaving a ${\rm \lsim 1\, M_\odot}$ carbon oxygen 
(C-O) white dwarf (WD) that cools slowly by radiation, possibly for 
billions of years. If the mass of such a C-O WD increases by accretion and 
approaches the Chandrasekhar mass limit, ${\rm M_{Ch}\approx 1.38M_\odot}$ 
(Chandrasekhar~1931), its core temperature rises and triggers a runaway 
thermonuclear explosion (Hoyle and Fowler 1960) in which no central object 
is left over. In such explosions, most of the nuclear binding energy 
release is converted to kinetic energy of the debris of type Ia supernova 
(SN), while the observed bolometric light-curve is powered mainly by the 
decay of the relatively long-lived end-product radioactive elements that 
were synthesized in the explosion (Colgate \& McKee 1969).

As the progenitors of core-collapse SNe are much more luminous than WDs, 
they have been identified from archival images on multiple occasions 
(e.g., Smartt 2009), while there have not yet been any direct observations 
of the WD progenitor of an SNIa. However, the SNeIa explosion paradigm is 
strongly supported by several observational facts (see, e.g., Hillebrandt, 
\&  Niemeyer 2000; Maoz \& Mannucci 2012; Astier 2012 
and references therein) such as:

 \begin{itemize} \item {WDs are produced by long-lived stars of less than 
${\rm\sim 8\, M_\odot}$, which eject most of their outer layers, leaving a 
${\rm\sim 1\, M_\odot}$ C-O WD that cools slowly by radiation, for 
billions of years.  Indeed SNeIa are the only type of SNe observed in old 
stellar environments such as elliptical galaxies.}

\item{The fast exothermic nuclear fusion reactions, starting with helium 
capture by $^{12}$C and $^{16}$O, produce the intermediate mass elements 
(IME) Si, S and Ca observed in the spectrum of SNeIa but not H and He, 
which  are lacking in the initial state and in the spectra of SNeIa.}
 
\item{The fast exothermic nuclear fusion reactions end with the near 
center production of $^{56}$Ni whose radioactive decay chain ${\rm 
^{56}Ni(\tau=8.76\, d)\rightarrow ^{56}Co(\tau=111.27\, d)\rightarrow 
^{56}Fe}$ seems to power the light-curves and explain the late appearance 
of iron group elements' spectral lines.}

\item{The kinetic energy of the debris, the bolometric 
light-curve, the spectrum and the spectral evolution of SNeIa 
are roughly those expected.}
\end{itemize}

Two main scenarios where the mass of a WD may approach ${\rm M_{c}}$ were 
proposed, the so called 'SD and DD scenarios'. In the single-degenerate 
(SD) scenario, the mass transfer is from a non-degenerate star (Whelan \& 
Iben 1973), while in the double degenerate (DD) scenario two WDs in a 
tight binary merge through loss of energy and angular momentum to 
gravitational waves (Iben \& Tutukov 1984, Webbink 1984). 
A third scenario where a WD may approach/cross ${\rm M_C}$ is 
WD collisions in triple star systems (Kusnir eta.~2013). But despite 
enormous observational efforts, it is still unknown by which mechanism the 
mass of a WD approaches  ${\rm M_C}$, which leads to its  
thermonuclear explosion. Furthermore, recent observations have put 
strong limits on the above  scenarios (see, e.g., Maoz and Mannucci 2012
Chomiuk 2013, and references therein). 

Another scenario in which WDs may reach a critical mass is mass accretion 
of fall-back matter by a nascent WDs in proto planetary nebulae. When the 
luminosity of the nascent WD decreases below the Eddington luminosity it 
may accrete fall-back He matter from the pulsation expulsion of the outer 
layers of the long-lived star of a mass less than ${\rm \sim 8\, 
M_\odot}$, until reaching a central thermonuclear ignition temperature. 
Moreover, the SNIa debris may collide with slowly moving circumstellar 
mass and produce a super luminous SNIa, such as SN 
2003fg (Howell et al.~2006) SN 2007if (Scalzo et al.~2010) and SN 2009dc 
(Silverman et al. 2011).

Because of the enormous diversity and complexity of the late phase of 
stellar evolution, mass expulsions and stellar explosions, it is natural 
to believe that only detailed numerical simulations with three dimensional 
hydrodynamics, thermonuclear energy release and transport by shocks, 
radiation and neutrinos are able to reproduce the observed light-curves 
and complex spectra of SNe. However, despite the complexity of SN 
explosions, SNeIa light-curves and spectra display an approximate 
'standard candle' behavior (see, e.g., Branch \& Tammann 1992) with 
simple correlations between various properties, such as peak luminosity 
and the decline rate of light-curves following the peak luminosity (e.g., 
Phillips~1993; Hamuy et al.~1996a,b,c,d,; Phillips~1999; Goobar and 
Perlmutter~1995; Riess et al.~1996; Tripp~1998). These empirical 
correlations were used to standardize SNe Ia and have allowed to improve 
the precision of cosmic distances estimated from SNeIa observations that 
led to the discovery of the accelerating expansion of the universe 
(Perlmutter et al.~1999; Riess et al.~1998).

The nearly standard properties of SNeIa during their photospheric (ph) 
phase suggest that perhaps the bolometric light-curves and other general 
properties of SNeIa can be obtained, to a good approximation, directly 
from a simple model, despite the complexity of SNeIa and the details of 
the complex radiation transport in their ejecta. Such semi-analytic 
approaches have been pioneered by Colgate and McKee~(1969), Arnett(1979), 
Colgate, Petschek and Kriese~(1980), Arnett~(1982). Improved semi-analytic 
solutions for the conversion of radioactive decay energy into the 
light-curves of SNe Ia have been proposed more recently, e.g., by Pinto 
\& Eastman~(2000) and Piro \& Nakar~(2013;2014). Here, using even a 
simpler analytical model, we derive the main properties of 
ordinary SNeIa, super Chandrasekhar SNeIa, and supeluminous SNeIa  
during their photospheric and early nebular phases,
which  depend on the $^{56}$Ni  mass produced in the thermouclear
explosion and the circumstellar mass. In particular,     
Collision with relatively slowly moving circumstellar material ejected 
before the SN explosion can enhance  
the bolometric light curve of ordinary SNeIa during their 
photospheric and nebular phase to appear as  
superluminous SNeIa or as an "ordinary" SNeIA powered by production 
of a Ni56 mass that exceeds the Chandrasekhar mass limit. 
In a following paper we also 
apply the master formula to describe other types of superluminous 
SNe such as SNeIb and super luminous SNeII 
(see, e.g., Gal-Yam~2012 and 
references therein for the current classification of SNe)

\section{Derivation of SNeIa General Properties}
We adopt the current paradigm of SNeIa explosions that 
mass accretion by a  carbon-oxygen 
(C-O) white dwarf that accretes fall-back matter or mass from a companion 
star and approaches a critical mass ${\rm M_C\approx M_{Ch}\approx 
1.38\, M_\odot}$, raises  its central 
temperature and triggers a thermonuclear explosion, which  produces IME 
via 
deflagration (Nomoto et al.~1976, 1984) followed by a transition to 
detonation that converts a fraction of the IME to $^{56}$Ni whose decay 
chain powers the bolometric light-curves of SNe Ia (Colgate \& McKee~1969; 
Arnett 1979; Colgate et al.~1980; Arnett 1982; Kuchner et al.~1994). The 
total mass of IME and iron 
group elements that is synthesized in the explosion, probably depends on 
unknown initial conditions, such as the distributions of density, 
composition and angular momentum in the WD progenitor when its mass 
approaches ${\rm M_C}$. For simplicity, we assume a spherical symmetry,
and that  
that the  amounts of $^{56}$Ni and IME that are synthesized in the SNIa 
explosion are roughly a constant fractions of ${\rm M_C}$.
synthesized during the SNIa explosion.

\subsection{The kinetic energy of the explosion:}  
The nuclear binding energy release per nucleon in the 
synthesis of C+O into typical IME,  such as 
$^{28}$Si and $^{40}$Ca (0.62 and 0.77 MeV respectively)
is not significantly  different from that released in the synthesis 
of $^{56}$Ni (0.815 MeV).
Only a small fraction of the nuclear  binding energy release
escapes by neutrino and photon emissions. Most of it 
is converted 
to the kinetic energy of the explosion. Moreover, when the mass of an
accreting WD crosses ${\rm M_C}$, the sum of the gravitational binding
energy and free energy of the degenerate electron gas  is $\approx 0$. 
Consequently, if  
kinetic energy ${\rm E_k}$ of the explosion 
is approximately the nuclear binding energy released in the synthesis of 
IME and  $^{56}$Ni. If approximately the entire mass 
${\rm M_C}$ is converted in the explosion to IME and $^{56}$Ni,
then,  
\begin{equation}
{\rm E_k\approx 
1.47\times 10^{51}\,(M_C/M_\odot)\, erg}\,.     
\label{Ek}
\end{equation}   
Hence, ${\rm E_k\approx 2 \times 10^{51}}$ ergs for ${\rm M_C\approx M_{Ch}}$.
                                         
\subsection{The expansion velocity} 
Early time spectroscopic observations 
of bright SNeIa show a bimodal expansion velocity (see, e.g., 
Childress et al. 2014). The high velocity component right after the 
explosion, as measured from the CA II IR triplet and Ca II H\&K and from 
Si II $\lambda$6355 is usually in the range between 20,000-30,000 km/s, 
while the lower velocity photospheric component as measured from the same 
lines and from the OI triplet and CII $\lambda$6580 and CII $\lambda$7234 
lines is usually in the range 14,000-16,000 km/s.

The bimodal photospheric velocity probably indicates a bimodal 
structure, such as an homologous expansion  plus a large number 
of higher velocity "bullets". 
Such structures were discovered in high resolution imaging of nearby young 
supernova remnants such as SNR Cas A (Fesen et al. 2006) and SNR 3C 58 
(Fesen et al. 2007) and in nearby planetary nebulae (PNe) such as the 
Helix nebula (Matsuura et al. 2007, 2009), and in many other PNe. The 
origin of 
these high velocity bullets is not clear. Probably they were expelled from 
the stellar surface by a Rayleigh-Taylor unstable shock/detonation-front 
propagating from the center of the star to its surface. In all the above 
cases the total mass and momentum of the "bullets" are a small fraction of 
the mass and momentum  of the exploding star.

The expansion velocity of the SN fireball can be estimated, assuming 
homologous expansion, i.e., a uniform spatial density throughout the
expanding mass at any moment. That implies that the expansion velocity 
${\rm v(r)}$ at distance ${\rm r}$ from the center satisfies 
${\rm v(r)=(r/R)\,V}$ where R 
is the radius of the fireball and ${\rm V=\dot{R}}$ is its radial 
expansion rate 
at R. Consequently, the total kinetic energy of the explosion is 
${\rm E_k=(3/10)\,M_C\,V^2} $ and the  initial expansion  velocity 
of SNeIa fireballs is ${\rm V_0 \approx 15,600\, km/s}$.

Note that for an homologous expansion, at Early time when the SN is 
highly opaque to radiation, i.e., when the optical depth of the SN fireball 
satisfies $\tau \gg 1$, the photospheric radius and photospheric velocity 
satisfy, respectively, ${\rm R_{ph}\approx R\,(1-2/3\,\tau)\approx R}$ and 
${\rm V_{ph}\approx V\,(1-2/3\, \tau)\approx V}$.

\subsection{The bolometric luminosity}
Let t be the time after shock break-out.
As long as  the SN fireball expands into "free space"  
its bolometric light-curve is powered mainly by trapping energy 
of gamma rays and positrons from the 
radioactive decay chain ${\rm ^{56}Ni\rightarrow ^{56}Co\rightarrow 
^{56}Fe}$.  Throughout the photospheric phase, the SN 
fireball is highly opaque to both optical photons and $\gamma$ rays. 
The thermal energy density 
${\rm u(T)}$ is dominated by black body radiation, i.e., 
${\rm u(T)\approx 7.56\times 10^{-15}\,T^4}$ ${\rm erg\, cm^{-3}\,K^{-4}}$,
and the total 
thermal energy of the SN fireball is ${\rm U\approx 4\pi\,R^3\,u(T)/3}$. 
During the photospheric phase, the SN fireball loses energy mainly by 
expansion, at a rate ${\rm \sim P\,dV/dt \approx U/t}$ (for a constant 
V), and 
by emission of photons, which are transported to the surface by a random 
walk. The photon emission yields a bolometric luminosity ${\rm L\approx 
U/t_{dif}}$, where ${\rm t_{dif}\approx R^2/\lambda\, c = R\, \tau/c}$ is 
the 
mean diffusion time of photons to the surface by a random walk, 
${\rm \lambda}$ 
is their mean free path and ${\rm \tau= R\,\Sigma\, n_i \sigma_i}$ is the 
fireball 
opacity where the summation extends over all particles in the fireball 
(ions, neutral atoms and free electrons) with density $n_i$ and effective 
cross section ${\rm \sigma_i}$. Roughly ${\rm n_i\sim R^{-3}}$ and ${\rm 
R=V\, t}$.  Hence, 
the mean diffusion time decreases with time roughly like 
${\rm t^{-1}}$ and can 
be written as ${\rm t_{dif}\approx{t_r^2/t}}$.  

During the photospheric phase, the opacity    
is mainly due to Compton scattering off free electrons, and hence
${\rm t_r \approx [3\,M_C\, f_e\,\sigma_{_T}/ 8\,\pi\,m_p\,c\,V]^{1/2}}$
where ${\rm \sigma_{_T}}$  is the Thomson cross section and
$f_e$ is the fraction of free (ionized)  electrons. Assuming that only
the 3s and 3p electrons outside the neon-like closed shells core of
IME such as Mg, Si, and S, and only the 4s and 3d electrons
outside the argon-like closed shell core of the iron group nuclei (IGN)
are ionized, one obtains ${\rm f_e(IME)\approx 
0.275}$ and  ${\rm f_e(IGN)=0.333}$, respectively. 
Thus, for ${\rm f_e\approx 0.30\pm 0.03}$ we 
expect ${\rm t_r\approx (11 \pm 1)\,(M_C/M_{Ch})^{1/2} d}$.

During the photospheric phase, energy conservation can be approximated by  
\begin{equation}
{\rm \dot{U}+U\,[{1\over t}+{1\over t_{dif}}]\approx \dot{E}}\,,
\label{Udeq}
\end{equation}
where ${\rm \dot{E}}$ is the energy deposition rate in the fireball
after the thermonuclear explosion by the decay of radioactive isotopes
synthesized in the explosion. 
The solution of Eq.~(2) is 
\begin{equation}
{\rm U={e^{-t^2/2\, t_r^2}\over t}\, 
\int_0^t t\,e^{t^2/2\, t_r^2} \dot{E}\,dt}\, .
\label{U}
\end{equation}
Consequently, the bolometric luminosity that satisfies 
${\rm L_b=t\,U/t_r^2}$
is given by the simple analytic expression
\begin{equation}
{\rm L_b={e^{-t^2/2\, t_r^2}\over t_r^2} 
\int_0^t t\,e^{t^2/2\, t_r^2} \dot{E}\, dt}.
\label{Lbol}                 
\end{equation}

For a luminosity that is powered  by  the radioactive decay chain  
${\rm ^{56}Ni\rightarrow ^{56}Co\rightarrow ^{56}Fe}$,
${\rm \dot{E}= \dot{E}_\gamma + \dot{E}_{e^+}}$ where    
\begin{equation}
{\rm \dot{E}_\gamma = {M(^{56}Ni)\over M_\odot}\,
[7.78\, A_\gamma(Ni)\,e^{-t/8.76\,d} + 1.50\,A_\gamma(Co)\, 
[e^{-t/111.27\,d}-e^{-t/8.76\,d}]]\,10^{43}\, erg\, s^{-1}}\,  
\label{Pgamma}
\end{equation}
is the power supply by $\gamma$-rays, and
\begin{equation}
{\rm \dot{E}_{e^+}= {M(^{56}Ni)\over 
M_\odot}\, A_e\,[e^{-t/111.27\,d}-e^{-t/8.76\,d}]\,10^{43}\, erg\, s^{-1}}\,. 
\label{betaplus} 
\end{equation} 
is the power supply by the kinetic energy 
loss of the positrons 
from the ${\rm \beta^+}$ decay of $^{56}$Co,
(branching ratio 19.48\%, average
positron kinetic energy 632.5 keV) which, presumably, are trapped by 
the turbulent magnetic field of the SN fireball. 
${\rm A_\gamma(Ni)}$ and  ${\rm A_\gamma(Co)}$ are the absorbed fractions 
of the$\gamma$-ray energy in the SN fireball
from the decay of $^{56}$Ni and $^{56}$Co, respectively. 
${\rm A_e\approx 0.05}$ is the ratio of the energy released as positron
kinetic energy and as $\gamma$-ray energy in the decay of $^{56}$Co
nuclei. 

For a uniformly distributed $^{56}$Ni over the entire SN fireball,
these absorbed fractions  are given roughly by
\begin{equation}
{\rm A_\gamma \approx 1-e^{-\tau_\gamma}},
\label{Agamma} 
\end{equation}
where 
\begin{equation}
{\rm \tau_\gamma={3\, M_C\,\sigma_t\over 8\,\pi\, m_p\, V^2\, t^2}=
 {t_\gamma^2\over t^2}}  
\label{tgamma}
\end{equation}
is the optical depth of the SN fireball, and ${\rm \sigma_t}$ is the 
effective cross section for energy transfer 
(${\rm dE_\gamma/dx=-\sigma_t\, n_e\, E_\gamma}$)   
to electrons  in Compton scattering.

The effective cross section for energy deposition in Compton scattering 
is obtained by integrating the 
Klein-Nishina (KN) energy transfer differential cross section over 
solid angle and by averaging over all the emitted $\gamma$ rays. 
In the KN domain, the average energy loss is a fraction 
${\rm \approx \epsilon /(1+2\,\epsilon)}$ of $E_\gamma$
and ${\rm \sigma_{KN}\approx 2.49\times 
10^{-25}\,(1+2\,ln\epsilon)/\epsilon\, cm^2}$
where ${\rm \epsilon=E_\gamma/m_e\, c^2}$.
The average $\gamma$-ray energies from the decay of $^{56}$Ni and 
$^{56}$Co are 0.53 MeV and 1.09 MeV, respectively. 
The corresponding effective energy transfer cross sections are ${\rm 
\sigma_t=9.5\times 
10^{-26}\, cm^2}$ and ${\rm 8.7\times 10^{-26}\,cm^2}$ 
for the  $^{56}$Ni and $^{56}$Co  $\gamma$-rays, 
respectively. 
They yield  ${\rm t_\gamma(Ni) \approx 33d }$ and ${\rm t_\gamma(Co) 
\approx 31d}$ for ${\rm M_C\approx M_{Ch}}$.
A single collision approximation is justified only when the SN 
fireball becomes semi-transparent  
to $\gamma$-rays (${\rm \tau_\gamma \lsim 1}$). 
A proper calculation of energy transfer in the multiple collisions
when ${\rm \tau_\gamma>1}$,  however, has 
only a small effect on ${\rm A_\gamma}$ and 
yields very similar bolometric light curves.

The positrons from the ${\rm \beta^+}$-decay of $^{56}$Co (and the ${\rm 
e^{\pm}}$ from the decay of other relatively long lived radioactive 
isotopes that were synthesized in the thermonuclear explosion) are 
presumably trapped in the SN fireball by its turbulent magnetic field. 
The ${\rm \beta^+}$ decay of $^{56}$Co and ${\rm \beta^{\pm}}$ 
from other long lived isotopes may dominate the power supply when the 
fireball becomes highly transparent to $\gamma$-rays and optical radiation 
during the nebular phase. During that phase, ionization and excitations by 
the $\gamma$-rays and $e^{\pm}$ lead to scintillation and bremsstrahlung 
emission, which dominate the SN emission.

\subsection{Early-time and late-time luminosities}   
${\rm \dot{E}}$ changes rather slowly with t 
relative to ${\rm t\,e^{t^2/2\, t_r^2}}$ 
and can be factored out of the integration in Eq.~(4), yielding
\begin{equation}
{\rm L_b\approx [1-e^{-t^2/2\, t_r^2}]\, \dot{E}}\, . 
\label{approxLb}
\end{equation}
Hence, the bolometric luminosity 
rises initially like ${\rm L_b\approx (t^2/2\,t_r^2)\,\dot{E}}$ and has 
the late-time asymptotic behavior ${\rm L_b\approx \dot{E}}$.
Note that the derivation of Eq.~(4) is valid only for the photospheric 
phase. However, its late-time behavior 
${\rm L_b(t)\approx \dot{E}(t)}$  is also the  correct behavior 
of ${\rm L_b}$ during the nebular phase. Thus Eq.~(9) is valid for both 
phases.

\subsection{The luminosity peak-time}
\noindent 
The approximate expression ${\rm L_b\sim [1-e^{-t^2/2\, 
t_r^2}]\,\dot{E}}$
peaks at ${\rm t=t_p\approx 17.5 \pm 1.5}$ d for 
${\rm t_r\approx 11\pm 1}$ d,
in good agreement with the  peak-time 
of ${\rm L_b(t)}$ given by Eq.~(4). The peak-time  depends on 
${\rm M_C}$  (${\rm t_r\propto M_C^{1/2}}$),
but not on  the synthesized mass of $^{56}$Ni.

\subsection{The peak luminosity - nickel mass relation for ordinary SNeIa}
\noindent
The peak value of the bolometric luminosity 
${\rm L_b(t)=t\,U/t_r^2}$ 
satisfies ${\rm \dot{L}_b=U/t_r^2 + t\,\dot{U}/t_r^2=0.}$
It then follows from Eq.~(4) that when the SN is still opaque to 
radiation at the peak time  ${\rm t=t_p}$, the peak luminosity 
satisfies ${\rm L_b(t_p)=\dot{E}(t_p)}$, which is  the Arnett relation 
(1979). In particular
for ${\rm t_p=17.5\pm 1.5}$ d,  the Arnett relation yields 
\begin{equation}
{\rm L_b(t_p)\approx (2.18\pm 0.17)\, \times 10^{43}\, {M(^{56}Ni)\over 
M_\odot}\,erg\, s^{-1}}. 
\label{Pgamma}
\end{equation}

\subsection{The color temperature during the photospheric phase}
As long as the fireball is optically thick (${\rm \tau\gg 1} $),
its continuum  spectrum is approximately that of a black body,
and its  luminosity is
given by the Stefan-Boltzmann law. For homologous expansion, the 
photospheric
velocity
decreases like ${\rm V_{ph}\approx V\, (1-2/3\,\tau))}$,
and the Stefan-Boltzmann law  (as long as $\tau \gg 1$) 
yields an effective photospheric temperature
\begin{equation}
{\rm T\approx \left[{[1-e^{-t^2/2\, t_r^2}]\, \dot{E}\over 4\,\pi\,
V_{ph}^2\,t^2\,\sigma}\right]^{1/4}}\,,
\label{Tc}
\end{equation}
where ${\rm \sigma=5.67\times 10^{-5}\, erg\, s^{-1}\, cm^{-2}\,
K^{-4}}$. During the transition
from the photospheric phase to the nebular phase, when free-free emission
and scintillations take over, the temperature decreases rather slowly.

\subsection{The  colour  light-curves and peak times}
As long as the SN fireball is optically thick, 
it radiates like Planck's  black body 
and the light-curves at a frequency $\nu$ satisfy
\begin{equation} 
{\rm L_\nu(t) = {8\,\pi^2\, R_{ph}^2\, h\, \nu^3\over c^2}\, {1\over  
e^{h\nu/kT}-1}}\,,   
\label{Lnu}
\end{equation}
where $h$ is the Planck constant, $k$ is the Boltzmann constant,  
${\rm R_{ph}=R\, (1-2\, V\, t^2/3\,c\,t_r^2)}$ and
${\rm V_{ph}=V\, (1-2\, V\, t^2/3\,c\,t_r^2)}$.
Near the peak-time of the bolometric luminosity,
the temperature has the approximate behavior
${\rm T(t)\approx T(t_p)\, (t_p/t)^{1/2}}$. Moreover, 
${\rm e^{h\nu/kT_p}\gg 1}$ in the  VBU bands, 
and since $R_{ph} \approx R\propto t$ for 
$t\lsim t_p$,
the maximum of ${L_\nu(t)}$ is reached  at a time  
\begin{equation}
{\rm t_{p,\nu}\approx 16 (kT_p/h\nu)^2\,t_p\propto 
\sqrt{M(^{56}Ni)}\, \nu^{-2}\,.}
\label{tpnu}
\end{equation}  

\subsection{The peak luminosity - decline rate correlation}
Although the peak intrinsic luminosity of SNeIa is not a standard 
candle, it appears to be correlated to the shape of their light-curves 
(Phillips~1993). Since 1993, various empirical correlations between the 
peak absolute magnitude of SNe Ia and the measured shapes of their 
intrinsic light-curves have been adopted for the use of SNe Ia as 
standard candles for distance measurements (e.g., Hamuy et 
al.~1996a,b,c,d; 
Riess et al.~1996, 1998; Perlmutter et al.~1999; Phillips et al. 1999; 
Goldhaber et al.~2001; Prieto et al.~2006). Most methods have used 
${\rm \Delta m_{15}}$, the magnitude difference in the intrinsic B-band 
light-curve  between maximum brightness and the brightness 15 days 
past it,  as a measure of the decline rate. 
To a good approximation, the B-band 
luminosity is proportional to the bolometric luminosity. 
Hence, ${\rm \Delta m_{15}(B)\approx 2.5\, 
log_{10}[L_b(t_p)/L_b(32.5d)]}$ where ${\rm L_b(t_p)}$ is given by 
Eq.~(10). 
At 32.5d, 
${\rm 1-e^{-t^2/ 2\,t_r^2}\approx 1}$, and ${\rm E_{e^+}\ll E_\gamma}$.
Thus, ${\rm L_b(32.5d)\approx \dot{E}(32.5d)\approx
A_\gamma(32.5d)\, \dot{E}_\gamma(32.5d)}$.
Consequently, it follows from Eqs.~(5) and (10) that 
for ${\rm\tau_\gamma\ll 1}$,
\begin{equation}
{\rm \Delta m_{15}(B)\approx 0.76 - log[A_\gamma(32.5d)]
\approx  0.76 - log[\tau_\gamma(32.5d)]}\,,
\label{dm15}
\end{equation}    
where ${\rm \tau_\gamma(t)=3\, M_C\,\sigma_t/8\,\pi\,m_p\,V_0^2\,t^2}\,.$
But,  ${\rm L_p \propto M(^{56}Ni)}$, 
and if  ${\rm M(^{56}Ni)\propto M_C}$, then roughly 
\begin{equation}
{\rm M_{max}\approx -M_0 +\Delta m_{15}(B)},
\label{PR}
\end{equation} 
which may explain, e.g., the correlation
${\rm M_{max}(B)=a + b(\Delta m_{15}(B)-1.1)}$ with
${\rm b=0.86\pm 0.21}$ (and ${\rm a=-19.256\pm 0.053}$) 
found by Hamuy et al.~(1996a,b,c,d)
for 18 ordinary SNeIa with a measured peak bolometric luminosity,  
assuming a Hubble constant ${\rm H_0=65\, km/s\, Mpc}$.

\section{Super luminous SNeIa}
SNeIa explosions can become super luminous 
by interaction with a circumstellar 
matter, such as a slowly expanding proto PN.
However, the unknown density distribution of the near circumstellar 
environment of SNeIa may be very complex.
For simplicity and demonstration purposes, consider a plastic collision 
(c) between a  fast expanding SNIa fireball with a velocity ${\rm V_c}$ 
and a slowly 
moving circumstellar (cs) spherical shell with a velocity ${\rm V_{cs}\ll 
V_c}$
that begins at ${\bf t_c}$  and ends at  ${\bf t_e}$. 
Assume that the cs has 
a wind-like  density profile,  
${\rm \rho(R)=\rho_0\ R_0^2/R^2}$ for ${\rm R_c\leq R\leq R_e}$
where ${\rm \rho_0\ R_0^2=\dot{M}/4\,\pi\,V_w}$, ${\rm V_{cs}=V_w}$.
and ${\rm M_{cs}=4\,\pi\, \rho_0\ R_0^2 \int_c^e V\,dt}$.
Neglecting  momentum loss through radiation and emission of cosmic 
ray particles,  
conservation of momentum during the collision  implies  that 
${\rm V=V_c\,M_C/M}$  where   ${\rm M=M_C+4\,\pi\, \rho_0\ R_0^2 
\int_c V\,dt}$, which yield 
\begin{equation}
{\rm {1\over V^3}\, {dV\over dt}=
-{4\,\pi\, \rho_0\, R_0^2\over M_C\, V_c}}.
\label{V(R)}
\end{equation}
Hence, the expansion velocity and radius as functionn of time are given by  
\begin{equation} 
{\rm V(t)=V_c/\sqrt{1+b\, (t-t_c)};\,\,\,\, 
R(t)=R_c+ 2\,(V_c /b)\, [\sqrt{1+b\,(t-t_c)}-1]}\,, 
\label{VR}
\end{equation}
respectively, where 
${\rm b= 8\,\pi\, \rho_0\ R_0^2\,V_c/ M_C}$. 
The swept in circumstellar mass by ${\rm M_C}$ is given by
${\rm M_{cs}\approx M_C\, (V_c/V(t_e)-1)}$.
The rate of 
mass loss by a 'constant wind' from the progenitor before the expolosion 
is  given by ${\rm \dot {M}_{cs}=b\, M_C\, V_w/2\, V_c}$ and  
the energy  deposition rate in the SN fireball by plasic collision with  
this massive 'wind' is 
\begin{equation}
{\rm \dot{E}_c(t)=  2\,\pi\, \rho_0\ R_0^2\,V^3  }.
\label{cP}
\end{equation}
This additional power supply must be included in ${\rm \dot E}$ in 
Eqs.~(2)-(4) as long as ${\rm t_c \leq t \leq t_{ec}}$.
During the collision, ${\rm t_{dif}= t_{rc}^2/t}$, and ${\rm t_{rc}}$ 
increases with time,
roughly like ${\rm t_{rc}=t_r(t_c)\,\sqrt{1+b(t-t_c)}}$.
After ${\rm t_{e}}$, when ${\rm \dot {E}_c(t)=0}$, Eq.~(2) yields
\begin{equation}
{\rm L_c(t>t_e)=L_c(t_e)\, 
exp^{-[(t-t_c)^2-(\Delta t_c)^2]/2\,t_{rc}^2}}
\label{Lctgte}
\end{equation}
where ${\rm  \Delta t_c=(t_e-t_c)}$ and 
${\rm t_{rc}(t_e)=t_{rc}(t_c)\, \sqrt{1+b\, \Delta t_c }}$.

\section{Comparison with observations} 
In figures 1-3 we compare our analytic expression Eq.~(4) and the observed 
rest-frame bolometric light-curves of three representative ordinary SNeIa 
(${\rm M_C=M_{Ch}}$) that were powered by the decay of $^{56}$Ni, had a 
very early detection and continuous follow up: SN 1992bc (Contardo et 
al.~2000), SN 1994ae (Contardo et al.~2000) and SN 2011fe in M101 (Nugent 
et al.~2011; Munari et al.~2012). In Figures 4-5 we compare the bolometric 
light-curve of the superluminous SN 2007if (Scalzo et al.~2010) and SN 
2009dc (Taubenberger et al.~2011) and our analytical expressions 
Eqs.~(4)-(8) and (17)-(19) for the bolometric light curve of superluminous 
SNeIa assuming they are powered by the decay of $^{56}$Ni and by collision 
with circumstellar matter. The values of the best fit parameters are 
listed in Table I.

Our best fits for the ordinary SNeIa, SN 2011fe (Nugent et al.~2011; 
Munari et al.~2012), SN 1994ae (Contardo et al.~2000) and SN 1992bc 
(Contardo et al.~2000), have yielded parameters consistent with their 
theoretical expectations for ${\rm M_C\approx M_{Ch}}$, namely ${\rm 
M(^{56}Ni)\lsim 0.5\, M_{Ch}}$, a negligible contribution, if any, from 
collision, and an ordinary expansion velocity.

The best fits obtained for the superluminous SNeIa explosions 
SN 2007if (Scalzo et al.~2010) and SN 2009dc (Taubenberger et al.~2011) 
have yielded a considerable  contribution from collision with a  
circumstellar matter with a much smaller  velocity (see 
Fig.~6), and  ${\rm M(^{56}Ni)< M_\odot}!$

For the best fit b=0.081/d and a typical proto PN radial expansion of 
$\sim 30$ km/s, the mass loss rate from the progenitor star before 
the SN explosion is ${\rm \dot{M}_{CS}\sim 0.044\, M_C\, y^{-1}}$.
The swept in circumstellar mass is 
${\rm M_{cs}\approx M_C\, (V_c/V(t_e)-1)}$.

\section{Conclusions} 
Supernovae type Ia, like all other types of supernovae, are a very complex 
astrophysical events that depend on the 
detailed late stage evolution of their progenitors   
and their environment. These 
unknowns determine in a complex way the properties of ordinary and 
superluminous SNeIa.  Nevertheless, here we have demonstrated that the 
observed bolometric lightcurves of normal  SNeIa during their 
photospheric and early nebular phase are well described by a simple 
analytic expression, which involves only five adjustable parameters 
(${\rm M(^{56}Ni),\, t_0, t_r,\,t_\gamma,\, A_e}$).
For sure, this is an over simplification of 
the diversity and complexity of SNIa explosions, and the 
demonstrated success 
of the master formula (Eq.~(4)) sto reproduce  their main properties is
partly due to the use of adjustable parameter. However, the fact that the 
values of these adjustable parameters are very close to their theoretical 
expected values, indicates that our simple model probably provides a 
useful simple description of the photospheric and early nebular 
phase of SNeIa.

Neither the energy deposition by positrons from the decay of Co56, nor 
the additional energy from recombination seems to be able to power the 
bolometric light curves of several SNe during their photospheric and 
nebular phase. In such cases circumstellar interaction may provide the 
additional power needed to explain their bolometric light curves during 
the photospheric and early nebular phase. Moreover, if the circumstellar 
interaction begins early enough, i.e., during the photospheric phase, it 
can supply a considerable fraction of the energy required to power 
superluminous SNeIa and delay the peak time of the bolometric light curve 
due to a larger copacity. In that case Eq.~(10)  is not valid.  If used, 
it overestimates the Ni56 mass synthesized in the explosion, and can even 
yield ${\rm M(Ni56)> M_{Ch}}$.

Superluminous SNeIa may be strongly interacting SNeIa whose bolometric 
light curve is powered by both the synthesis of of $^{56}$Ni and early
collision with circumstellar matter such as that around the center of 
proto-planetary nebulae. This has been demonstrated in this paper , 
admittedly, using an 
oversimplified model and a couple of adjustable parameters. 
Interestingly, the model best 
fits of the bolometric lightcurve of the superluinous SN 2007if and SN 
2009dc yield ${\rm M(Ni56)\approx 0.78\, M_\odot}$ and ${\rm 
M(Ni56)\approx 0.90\, M_\odot}$, respectively. These values are within the 
observed range of the values of ${\rm M(Ni56)}$ obtained for ordinary 
SNeIa.

A significant fraction of ordinary SNeIa could take place during PN 
formation or a failed PN formation: The X-ray and radio upper limits of 
nearby, normal type Ia, which show no sign of circumstellar gas to very 
faint limits may be explained if the fall back takes place at an early 
stage of the PN (or "failed"  PN)  formation. However, it still remains to 
be tested both by detailed numerical calculations (e.g., G. Shaviv et al. 
to be published) and by conclusive spectroscopic observations whether 
accretion of fall-back matter (mainly He4) onto a  nascent WDs at the 
center of proto-PN can trigger a significant fraction if not most of the 
ordinary SNeIa and superluminous SNeIa.

{\bf Acknowledgement:} We thank an anonymous referee for very
careful examination of our manuscript, and for constructive criticism  
and uggestions.

\begin{deluxetable}{llllllllll}
\tablewidth{0pt}
\tablecaption{Best fit parameters of the 
analytic description of the bolometric light-curves 
of the SNeIa shown in Figs.~1-5. ${\rm t_0}$ is the explosion time 
relative to maximum light}
\tablehead{
\colhead{SNIa} &\colhead{${\rm t_0}$ [d]} & \colhead{${\rm t_r}$ [d]} &
\colhead{${\rm t_\gamma}$ [d]} & \colhead{M($^{56}$Ni)} 
&\colhead{${\rm A_e}$}&  &   &    }
\startdata
SN 1992bc   & -19.6 & 12.6 & 26.8 & 0.84${\rm M_\odot}$  &~~~0.15& & \\
SN 1994ae   & -16.1 &  10.0 & 28.2 & 0.47${\rm M_\odot}$ &~~~0.18& & \\
SN 2011fe  & -17.5 & 14.6 & 19.5 & 0.73${\rm M_\odot}$   &~~~0.12& & \\ 
\hline
average & -17.7 & 12.4 & $24.8$ & 0.68${\rm M_\odot}$   &~~~0.15& & \\
theory$(\approx)$  &-17.5 & 11.5 & 29.0 & ${\rm <M_{CH}}$& $~>0.06$&& \\
\hline
SLSNIa & & & & &${\rm t_c}$ [d]& ${\rm t_{rc}}$ [d] & b/[d] & ${\rm\Delta 
t_c}$[d]\\
\hline
SN 2007if  & -21.3 & 11.5 & 36.1 &  0.78 $M_\odot$& -2.93 &17.3 &~0.081 
&~~58.3\\
SN 2009dc  & -21.4 & 12.5 & 25.9 &  0.92 $M_\odot$&-5.5& 16.6 
&~0.085&~~42.3\\
\hline
\enddata
\end{deluxetable}

\newpage 
\begin{figure}[]
\centering 
\vspace{-2cm} 
\epsfig{file=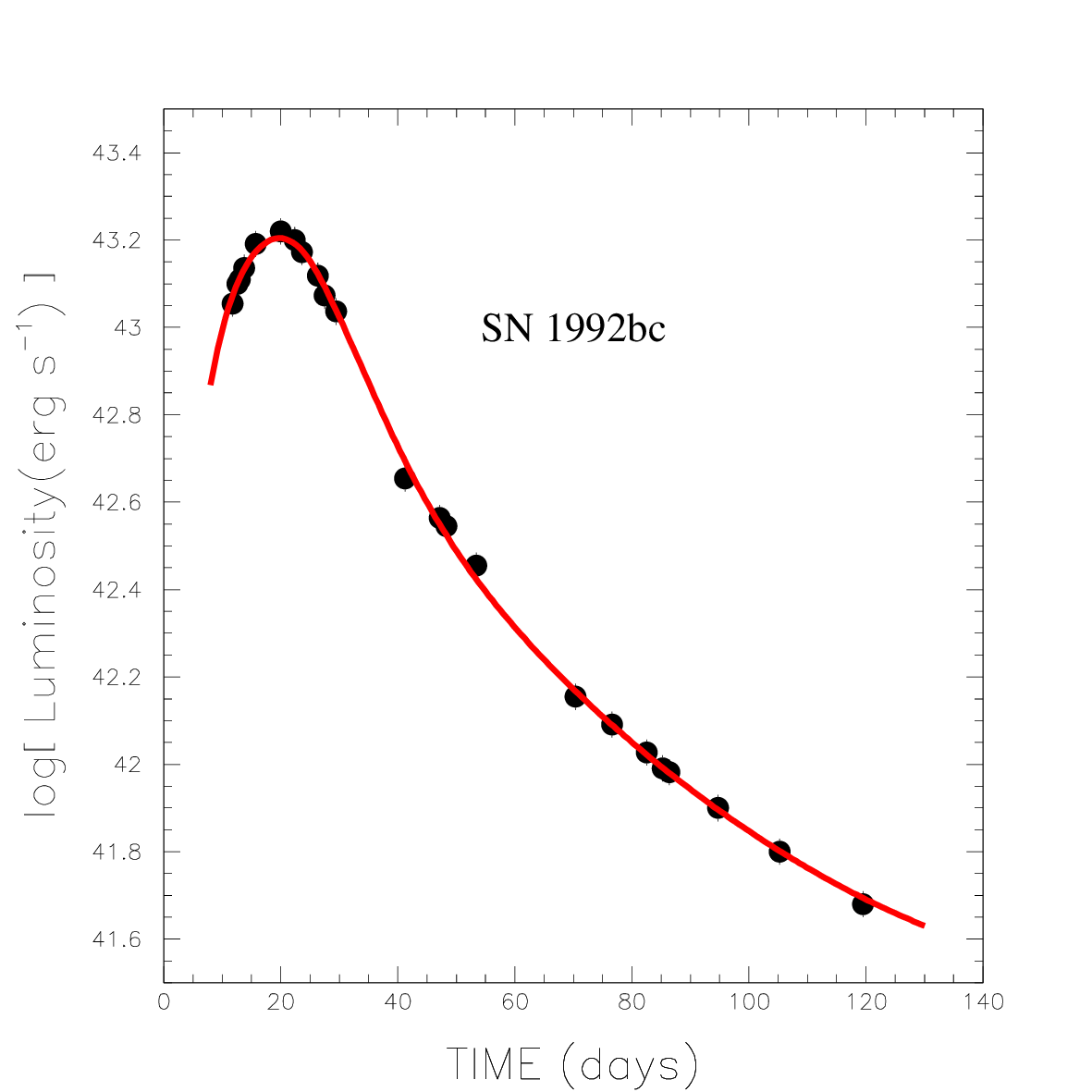,width=12.cm,height=12.cm} 
\caption{ 
Comparison between the bolometric light-curve of SN 1992bc 
(Contardo et al.~2000) and that predicted by the
analytic model and summarized in Eqs.~(4)-(8),
assuming it was powered by the decay  of $^{56}$Ni, 
synthesized in the thermonuclear explosion of a C-O white dwarf of 
a critical mass ${\rm M_C=M_{Ch}}$.}
\label{Fig1}
\end{figure}

\newpage 
\begin{figure}[]
\centering 
\vspace{-2cm} 
\epsfig{file=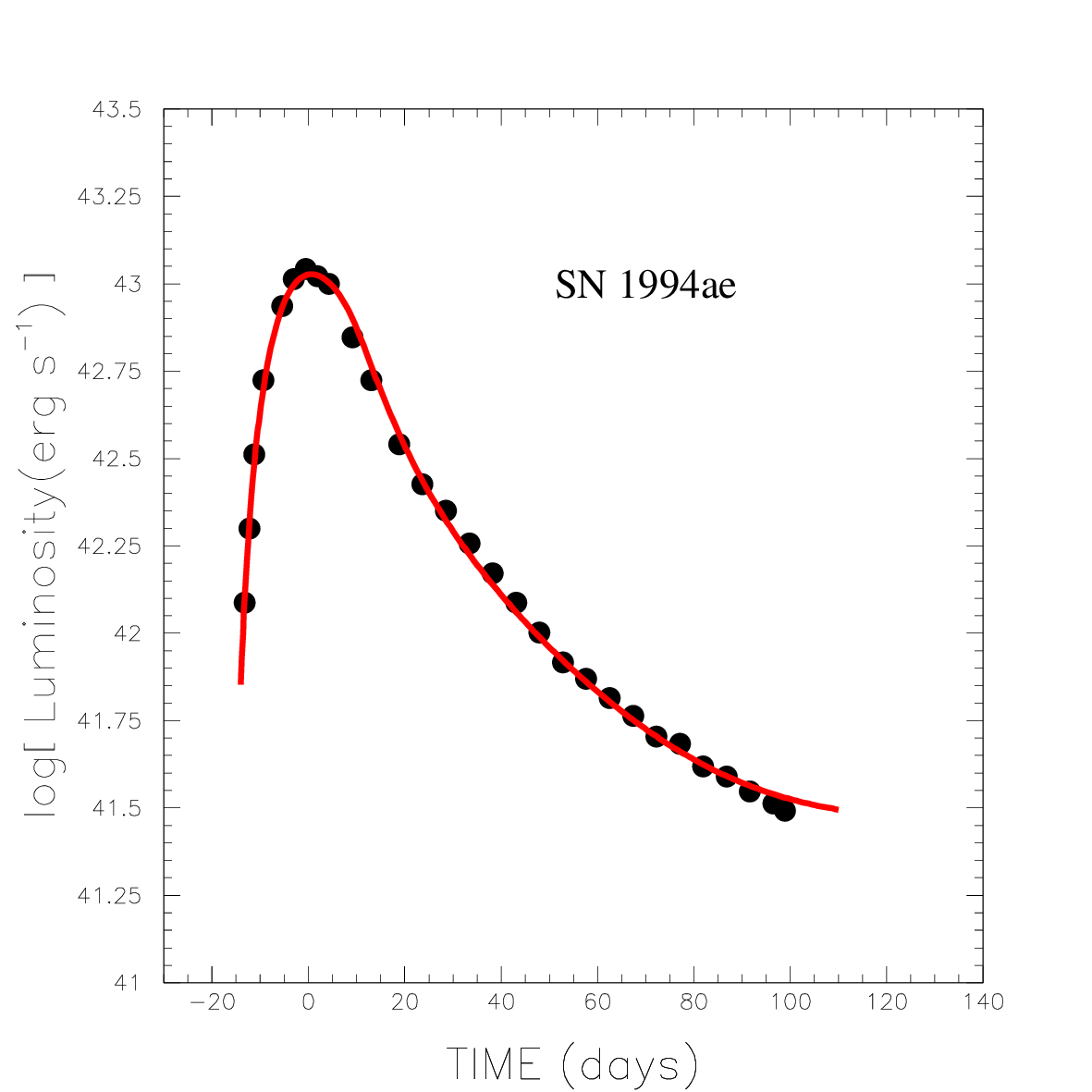,width=12.cm,height=12.cm} 
\caption{Comparison between the measured bolometric light-curve of 
SN 1994ae (Contardo et al.~2000) and that predicted by the analytic 
model and summarized in Eqs.~(4)-(8),
assuming it was powered by the decay  of $^{56}$Ni,
synthesized in the thermonuclear explosion of a C-O white dwarf of
a critical mass ${\rm M_C=M_{Ch}}$.}
\label{Fig2}
\end{figure}

\newpage
\begin{figure}[]
\centering
\vspace{-2cm}
\epsfig{file=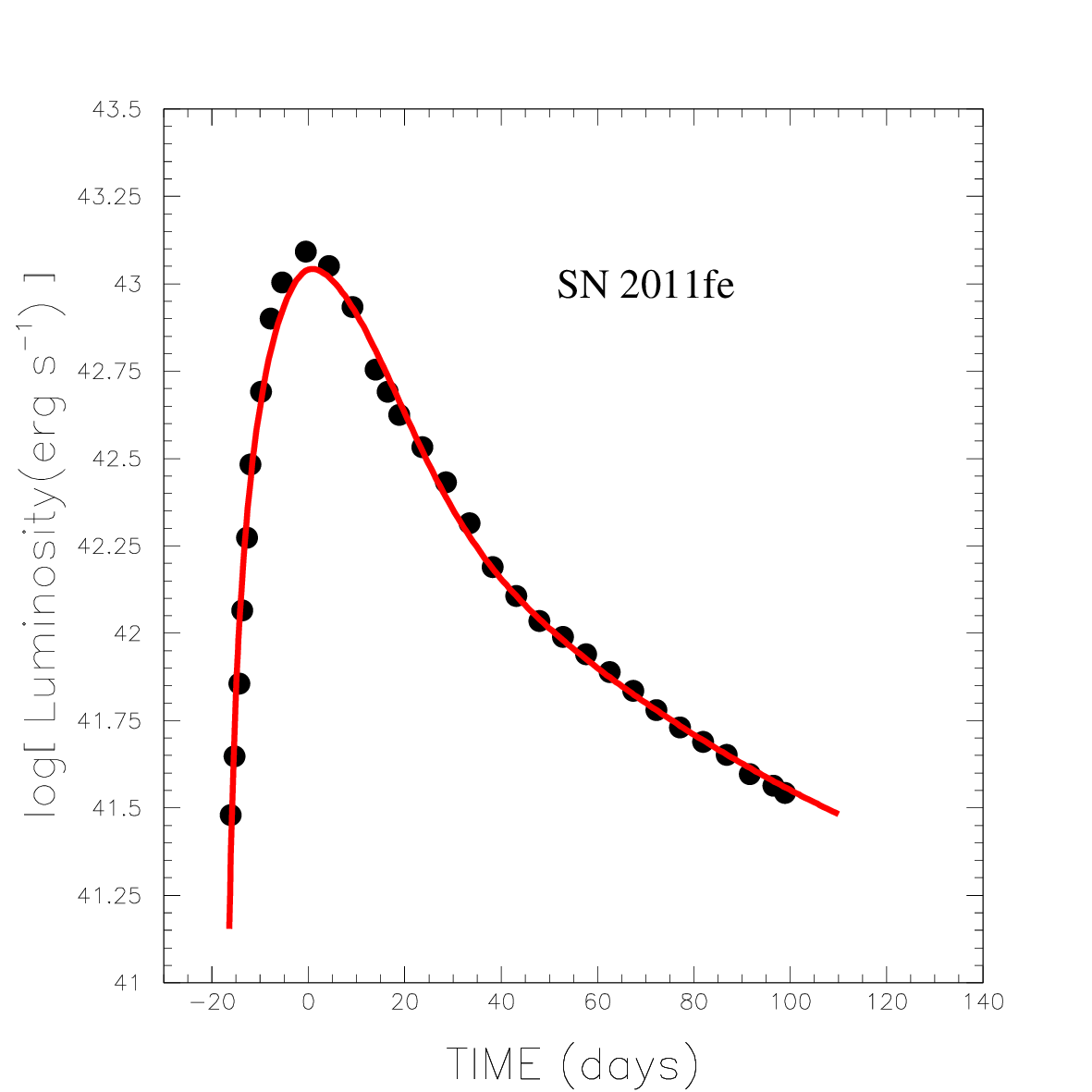,width=12.cm,height=12.cm}
\caption{
Comparison between the bolometric light-curve of SN 2011fe 
(Munari et al.~2012) and that predicted by the analytic model 
and summarized in Eqs.~(4)-(8),
assuming it was powered by the decay  of $^{56}$Ni,
synthesized in the thermonuclear explosion of a C-O white dwarf of
a critical mass ${\rm M_C=M_{Ch}}$.}
\label{Fig3}
\end{figure}

\newpage 
\begin{figure}[]
\centering 
\vspace{-2cm} 
\epsfig{file=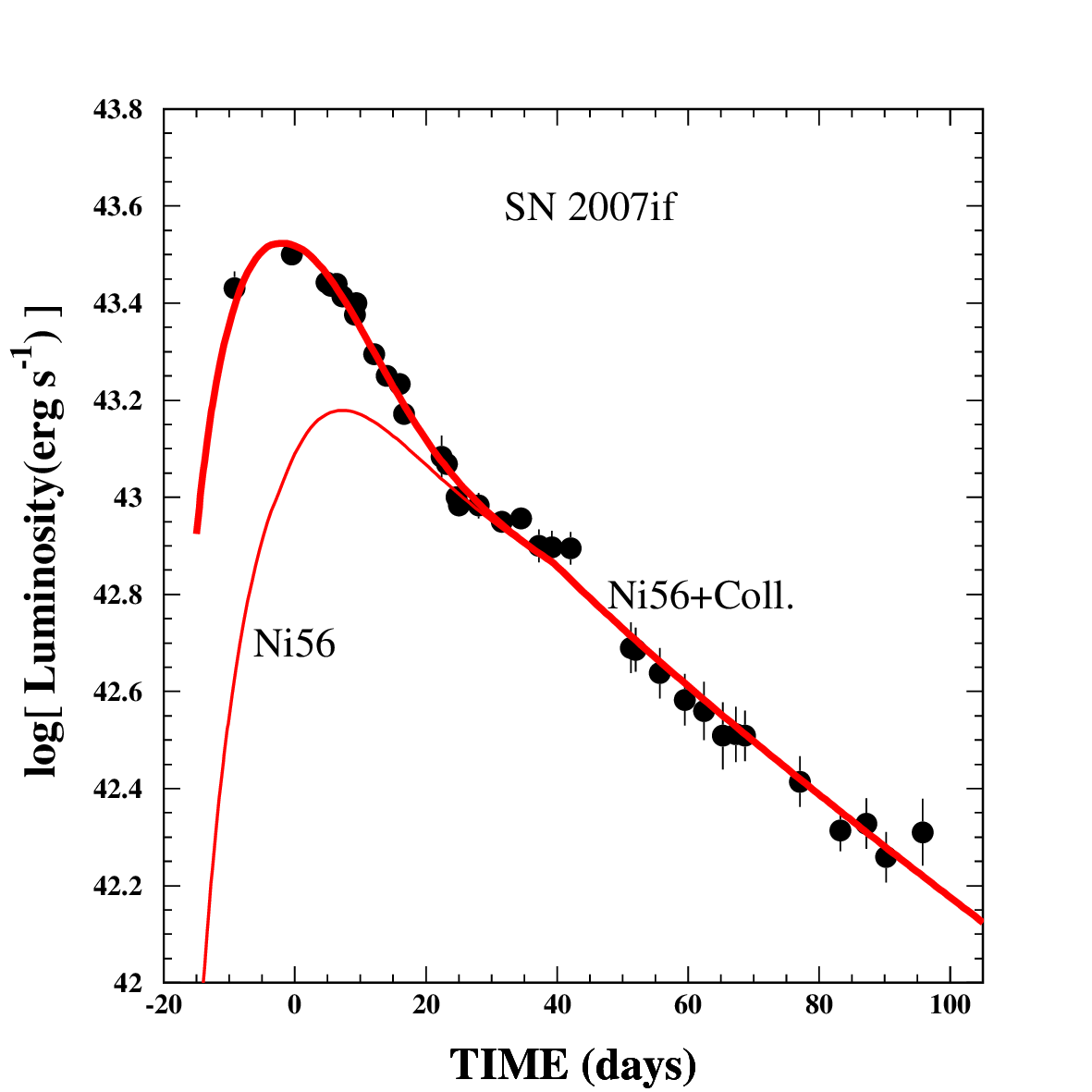,width=12.cm,height=12.cm} 
\caption{ 
Comparison between the bolometric light-curve of the superluminous 
SN 2007if (Scalzo et al.~2010) and that predicted by the analytical model 
(thick line) as summarized by Eqs. (4)-(8) and (17)-(18)  
assuming it was powered by both  
the decay  of $^{56}$Ni (thin line) and the collision 
with a fall-back circumstellar matter.}
\label{Fig4}
\end{figure}

\newpage
\begin{figure}[] 
\centering 
\vspace{-2cm} 
\epsfig{file=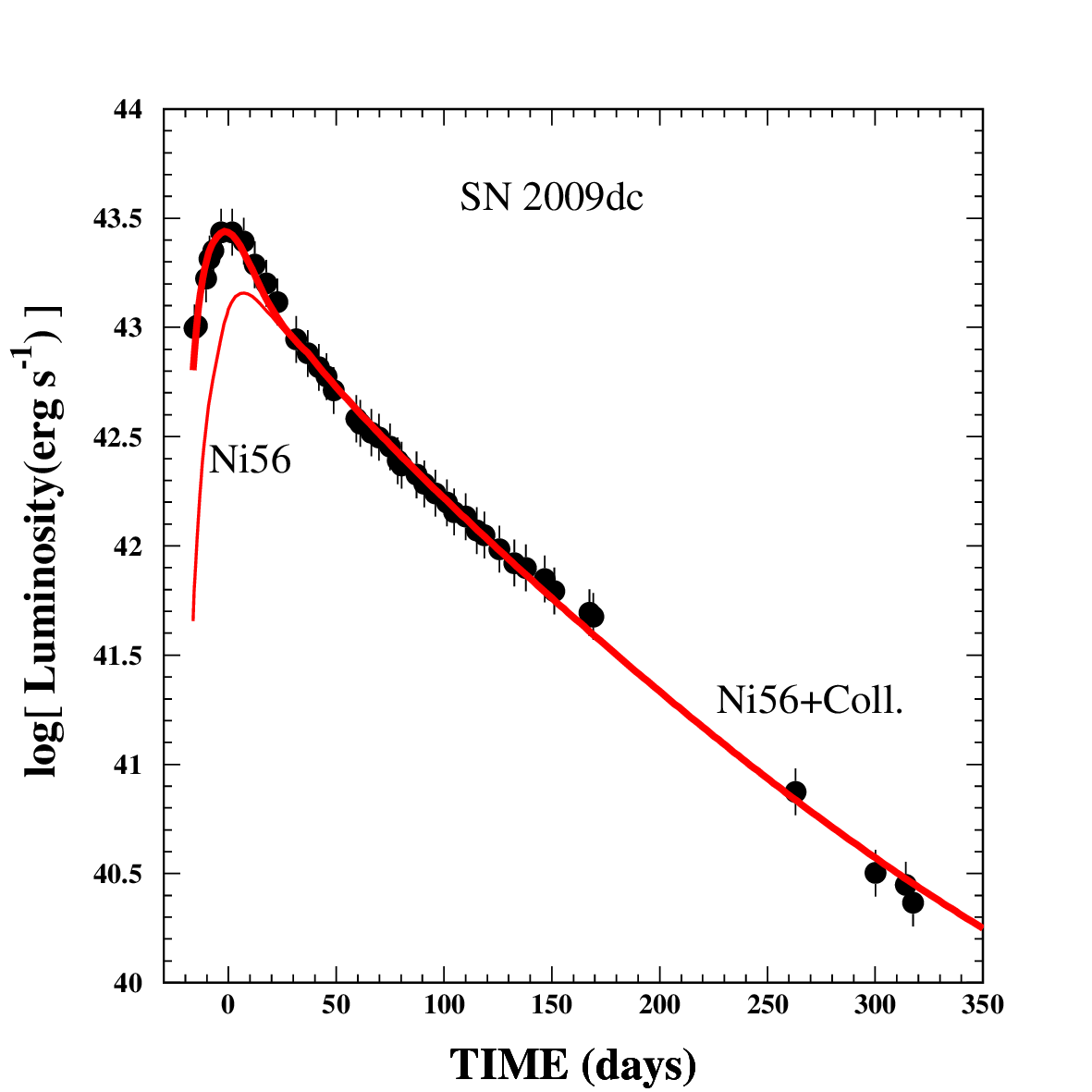,width=12.cm,height=12.cm} 
\caption{
Comparison between the bolometric light-curve of the superluminous
SN 2009dc (Taubenberger et al.~2011) and that predicted by 
the analytical model
(thick line) as summarized by Eqs. (4)-(8) and (17)-(18) 
assuming it was powered by both
the decay  of $^{56}$Ni (thin line) and the collision
with a fall-back circumstellar matter.}

\label{Fig5} 
\end{figure}

\newpage
\begin{figure}[]
\centering   
\vspace{-2cm}
\epsfig{file=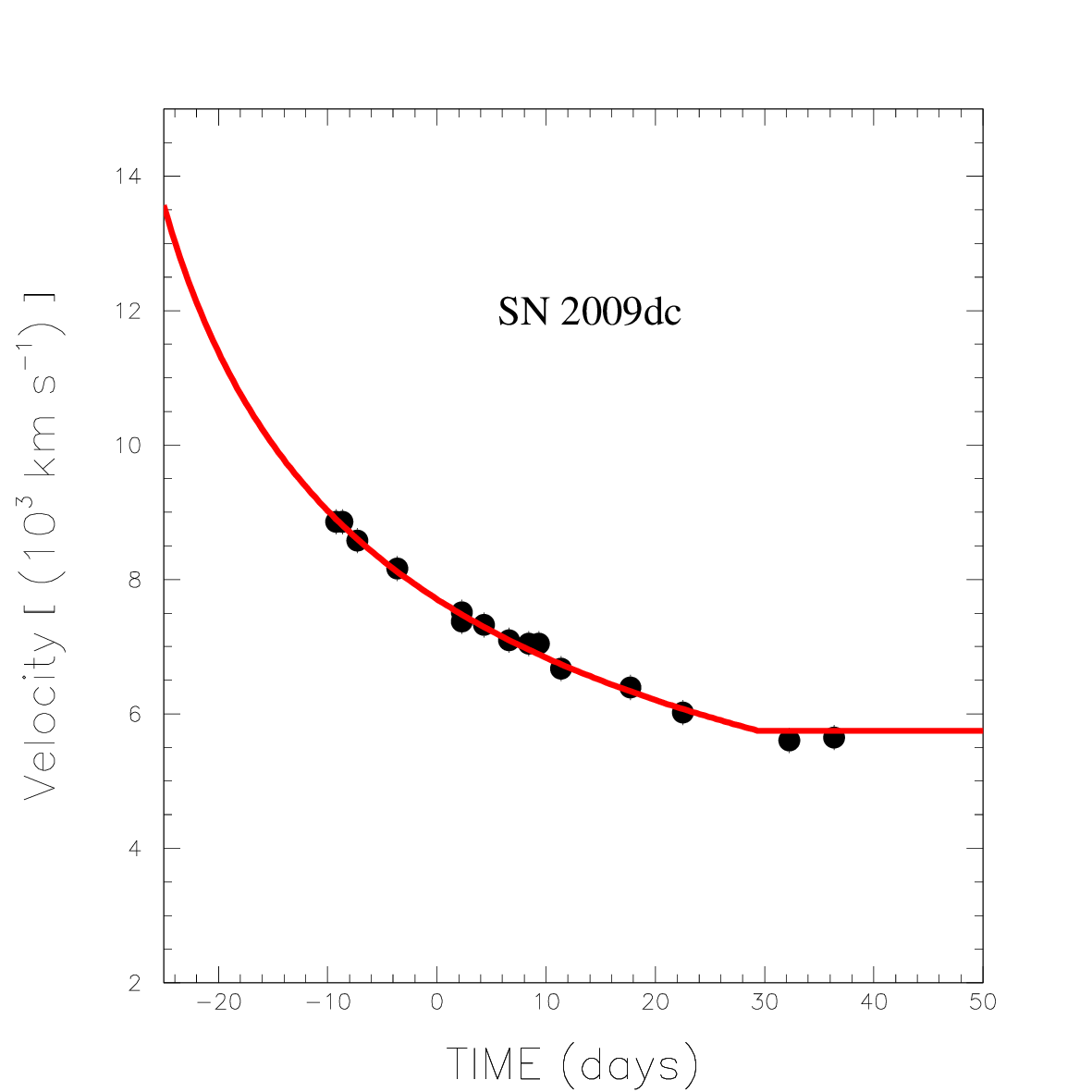,width=12.cm,height=12.cm}
\caption{
Comparison between the spectroscopically inferred decline of the 
expansion velocity of the superluminous
SN 2009dc (Taubenberge et al.~2011) and that obtained from the best fit to
its  bolometric 
light curve shown in Fig.~5, 
assuming it was powered by the decay  of $^{56}$Ni,
synthesized in the thermonuclear explosion of a  white dwarf 
within a fall-back circumstellar matter.}
\label{Fig6}
\end{figure}

\end{document}